

%
\newbox\hdbox%
\newcount\hdrows%
\newcount\multispancount%
\newcount\ncase%
\newcount\ncols
\newcount\nrows%
\newcount\nspan%
\newcount\ntemp%
\newdimen\hdsize%
\newdimen\newhdsize%
\newdimen\parasize%
\newdimen\spreadwidth%
\newdimen\thicksize%
\newdimen\thinsize%
\newdimen\tablewidth%
\newif\ifcentertables%
\newif\ifendsize%
\newif\iffirstrow%
\newif\iftableinfo%
\newtoks\dbt%
\newtoks\hdtks%
\newtoks\savetks%
\newtoks\tableLETtokens%
\newtoks\tabletokens%
\newtoks\widthspec%
%
%
\immediate\write15{%
CP SMSG GJMSINK TEXTABLE --> TABLE MACROS V. 851121 JOB = \jobname%
}%
%
%
\tableinfotrue%
\catcode`\@=11
%
%
\def\tstrut{\vrule height3.1ex depth1.2ex width0pt}%
\def\and{\char`\&}
\def\tablerule{\noalign{\hrule height\thinsize depth0pt}}%
\thicksize=1.5pt
\thinsize=0.6pt
\def\thickrule{\noalign{\hrule height\thicksize depth0pt}}%
\def\ctr#1{\hfil\ #1\hfil}%
%
%
%
%
\tablewidth=-\maxdimen%
\spreadwidth=-\maxdimen%
\def\tabskipglue{0pt plus 1fil minus 1fil}%
%
%
\centertablestrue%
%
%
%
%
\parasize=4in%
\gdef\ARGS{########}
\gdef\headerARGS{####}
\def\@mpersand{&}
{\catcode`\|=13
\gdef\letbarzero{\let|0}
\gdef\letbartab{\def|{&&}}%
\gdef\letvbbar{\let\vb|}%
}
{\catcode`\&=4
\def\ampskip{&\omit\hfil&}
\catcode`\&=13
\let&0
\xdef\letampskip{\def&{\ampskip}}%
\gdef\letnovbamp{\let\novb&\let\tab&}
}
\def\begintable{
   \begingroup%
   \catcode`\|=13\letbartab\letvbbar%
   \catcode`\&=13\letampskip\letnovbamp%
   \def\multispan##1{
      \mscount##1%
      \multiply\mscount\tw@\advance\mscount\m@ne%
      \loop\ifnum\mscount>\@ne \sp@n\repeat%
   }
   \def\|{%
      &\omit\widevline&%
   }%
   \ruledtable
}
\long\def\ruledtable#1\endtable{%
%
%
%
   \offinterlineskip
   \tabskip 0pt
   \def\widevline{\vrule width\thicksize}
   \def\endrow{\@mpersand\omit\hfil\crnorm\@mpersand}%
   \def\crthick{\@mpersand\crnorm\thickrule\@mpersand}%
   \def\crthickneg##1{\@mpersand\crnorm\thickrule
          \noalign{{\skip0=##1\vskip-\skip0}}\@mpersand}%
   \def\crnorule{\@mpersand\crnorm\@mpersand}%
   \def\crnoruleneg##1{\@mpersand\crnorm
          \noalign{{\skip0=##1\vskip-\skip0}}\@mpersand}%
   \let\nr=\crnorule
   \def\endtable{\@mpersand\crnorm\thickrule}%
   \let\crnorm=\cr
%
%
   \edef\cr{\@mpersand\crnorm\tablerule\@mpersand}%
   \def\crneg##1{\@mpersand\crnorm\tablerule
          \noalign{{\skip0=##1\vskip-\skip0}}\@mpersand}%
   \let\ctneg=\crthickneg
   \let\nrneg=\crnoruleneg
   \the\tableLETtokens
%
%
   \tabletokens={&#1}
%
%
   \countROWS\tabletokens\into\nrows%
   \countCOLS\tabletokens\into\ncols%
%
%
   \advance\ncols by -1%
   \divide\ncols by 2%
   \advance\nrows by 1%
%
%
   \iftableinfo %
      \immediate\write16{[Nrows=\the\nrows, Ncols=\the\ncols]}%
   \fi%
%
%
   \ifcentertables
      \ifhmode \par\fi
      \line{
      \hss
   \else %
      \hbox{%
   \fi
      \vbox{%
         \makePREAMBLE{\the\ncols}
         \edef\next{\preamble}
         \let\preamble=\next
         \makeTABLE{\preamble}{\tabletokens}
      }
      \ifcentertables \hss}\else }\fi
   \endgroup
   \tablewidth=-\maxdimen
   \spreadwidth=-\maxdimen
}
\def\makeTABLE#1#2{
   {
   \let\ifmath0
   \let\header0
   \let\multispan0
%
%
   \ncase=0%
   \ifdim\tablewidth>-\maxdimen \ncase=1\fi%
   \ifdim\spreadwidth>-\maxdimen \ncase=2\fi%
   \relax
%
   \ifcase\ncase %
      \widthspec={}%
   \or %
      \widthspec=\expandafter{\expandafter t\expandafter o%
                 \the\tablewidth}%
   \else %
      \widthspec=\expandafter{\expandafter s\expandafter p\expandafter r%
                 \expandafter e\expandafter a\expandafter d%
                 \the\spreadwidth}%
   \fi %
   \xdef\next{
      \halign\the\widthspec{%
      #1
      \noalign{\hrule height\thicksize depth0pt}
      \the#2\endtable
%
      }
   }
   }
   \next
}
\def\makePREAMBLE#1{
   \ncols=#1
   \begingroup
   \let\ARGS=0
   \edef\xtp{\widevline\ARGS\tabskip\tabskipglue%
   &\ctr{\ARGS}\tstrut}
   \advance\ncols by -1
   \loop
      \ifnum\ncols>0 %
      \advance\ncols by -1%
      \edef\xtp{\xtp&\vrule width\thinsize\ARGS&\ctr{\ARGS}}%
   \repeat
   \xdef\preamble{\xtp&\widevline\ARGS\tabskip0pt%
   \crnorm}
   \endgroup
}
\def\countROWS#1\into#2{
   \let\countREGISTER=#2%
   \countREGISTER=0%
   \expandafter\ROWcount\the#1\endcount%
}%
\def\ROWcount{%
   \afterassignment\subROWcount\let\next= %
}%
\def\subROWcount{%
   \ifx\next\endcount %
      \let\next=\relax%
   \else%
      \ncase=0%
      \ifx\next\cr %
         \global\advance\countREGISTER by 1%
         \ncase=0%
      \fi%
      \ifx\next\endrow %
         \global\advance\countREGISTER by 1%
         \ncase=0%
      \fi%
      \ifx\next\crthick %
         \global\advance\countREGISTER by 1%
         \ncase=0%
      \fi%
      \ifx\next\crnorule %
         \global\advance\countREGISTER by 1%
         \ncase=0%
      \fi%
      \ifx\next\crthickneg %
         \global\advance\countREGISTER by 1%
         \ncase=0%
      \fi%
      \ifx\next\crnoruleneg %
         \global\advance\countREGISTER by 1%
         \ncase=0%
      \fi%
      \ifx\next\crneg %
         \global\advance\countREGISTER by 1%
         \ncase=0%
      \fi%
      \ifx\next\header %
         \ncase=1%
      \fi%
      \relax%
      \ifcase\ncase %
         \let\next\ROWcount%
      \or %
         \let\next\argROWskip%
      \else %
      \fi%
   \fi%
   \next%
}
\def\counthdROWS#1\into#2{%
\dvr{10}%
   \let\countREGISTER=#2%
   \countREGISTER=0%
\dvr{11}%
\dvr{13}%
   \expandafter\hdROWcount\the#1\endcount%
\dvr{12}%
}%
\def\hdROWcount{%
   \afterassignment\subhdROWcount\let\next= %
}%
\def\subhdROWcount{%
   \ifx\next\endcount %
      \let\next=\relax%
   \else%
      \ncase=0%
      \ifx\next\cr %
         \global\advance\countREGISTER by 1%
         \ncase=0%
      \fi%
      \ifx\next\endrow %
         \global\advance\countREGISTER by 1%
         \ncase=0%
      \fi%
      \ifx\next\crthick %
         \global\advance\countREGISTER by 1%
         \ncase=0%
      \fi%
      \ifx\next\crnorule %
         \global\advance\countREGISTER by 1%
         \ncase=0%
      \fi%
      \ifx\next\header %
         \ncase=1%
      \fi%
\relax%
      \ifcase\ncase %
         \let\next\hdROWcount%
      \or%
         \let\next\arghdROWskip%
      \else %
      \fi%
   \fi%
   \next%
}%
{\catcode`\|=13\letbartab
\gdef\countCOLS#1\into#2{%
   \let\countREGISTER=#2%
   \global\countREGISTER=0%
   \global\multispancount=0%
   \global\firstrowtrue
   \expandafter\COLcount\the#1\endcount%
   \global\advance\countREGISTER by 3%
   \global\advance\countREGISTER by -\multispancount
}%
\gdef\COLcount{%
   \afterassignment\subCOLcount\let\next= %
}%
{\catcode`\&=13%
\gdef\subCOLcount{%
   \ifx\next\endcount %
      \let\next=\relax%
   \else%
      \ncase=0%
      \iffirstrow
         \ifx\next& %
            \global\advance\countREGISTER by 2%
            \ncase=0%
         \fi%
         \ifx\next\span %
            \global\advance\countREGISTER by 1%
            \ncase=0%
         \fi%
         \ifx\next| %
            \global\advance\countREGISTER by 2%
            \ncase=0%
         \fi
         \ifx\next\|
            \global\advance\countREGISTER by 2%
            \ncase=0%
         \fi
         \ifx\next\multispan
            \ncase=1%
            \global\advance\multispancount by 1%
         \fi
         \ifx\next\header
            \ncase=2%
         \fi
         \ifx\next\cr       \global\firstrowfalse \fi
         \ifx\next\endrow   \global\firstrowfalse \fi
         \ifx\next\crthick  \global\firstrowfalse \fi
         \ifx\next\crnorule \global\firstrowfalse \fi
         \ifx\next\crnoruleneg \global\firstrowfalse \fi
         \ifx\next\crthickneg  \global\firstrowfalse \fi
         \ifx\next\crneg       \global\firstrowfalse \fi
      \fi
\relax
      \ifcase\ncase %
         \let\next\COLcount%
      \or %
         \let\next\spancount%
      \or %
         \let\next\argCOLskip%
      \else %
      \fi %
   \fi%
   \next%
}%
\gdef\argROWskip#1{%
   \let\next\ROWcount \next%
}
\gdef\arghdROWskip#1{%
   \let\next\ROWcount \next%
}
\gdef\argCOLskip#1{%
   \let\next\COLcount \next%
}
}
}
\def\spancount#1{
   \nspan=#1\multiply\nspan by 2\advance\nspan by -1%
   \global\advance \countREGISTER by \nspan
   \let\next\COLcount \next}%
\def\dvr#1{\relax}%
\def\header#1{%
\dvr{1}{\let\cr=\@mpersand%
\hdtks={#1}%
\counthdROWS\hdtks\into\hdrows%
\advance\hdrows by 1%
\ifnum\hdrows=0 \hdrows=1 \fi%
\dvr{5}\makehdPREAMBLE{\the\hdrows}%
\dvr{6}\getHDdimen{#1}%
{\parindent=0pt\hsize=\hdsize{\let\ifmath0%
\xdef\next{\valign{\headerpreamble #1\crnorm}}}\dvr{7}\next\dvr{8}%
}%
}\dvr{2}}
\def\makehdPREAMBLE#1{
\dvr{3}%
\hdrows=#1
{
\let\headerARGS=0%
\let\cr=\crnorm%
\edef\xtp{\vfil\hfil\hbox{\headerARGS}\hfil\vfil}%
\advance\hdrows by -1
\loop
\ifnum\hdrows>0%
\advance\hdrows by -1%
\edef\xtp{\xtp&\vfil\hfil\hbox{\headerARGS}\hfil\vfil}%
\repeat%
\xdef\headerpreamble{\xtp\crcr}%
}
\dvr{4}}
\def\getHDdimen#1{%
\hdsize=0pt%
\getsize#1\cr\end\cr%
}
\def\getsize#1\cr{%
\endsizefalse\savetks={#1}%
\expandafter\lookend\the\savetks\cr%
\relax \ifendsize \let\next\relax \else%
\setbox\hdbox=\hbox{#1}\newhdsize=1.0\wd\hdbox%
\ifdim\newhdsize>\hdsize \hdsize=\newhdsize \fi%
\let\next\getsize \fi%
\next%
}%
\def\lookend{\afterassignment\sublookend\let\looknext= }%
\def\sublookend{\relax%
\ifx\looknext\cr %
\let\looknext\relax \else %
   \relax
   \ifx\looknext\end \global\endsizetrue \fi%
   \let\looknext=\lookend%
    \fi \looknext%
}%
%
%
\def\tablelet#1{%
   \tableLETtokens=\expandafter{\the\tableLETtokens #1}%
}%
\catcode`\@=12
%
\input jnl
\input reforder
\hsize=15.5cm\vsize=23.5cm\hoffset=.5cm\voffset=0cm
\input tables2
\def\ccr{\cr\noalign{\vskip 0.2cm}}
\nopagenumbers
\singlespace
\vskip 2.5cm

\centerline{{\twelvepoint\bf {SYMMETRIC TEXTURES }}
\footnote{*}{Invited Talk at Global Foundation Conference, Coral Gables, Jan
1993}}


\vskip 1.1cm
\centerline{Pierre Ramond
\footnote{**}{Work supported in part by the Department of Energy,
contract DE-FG05-86ER-40272} }

\vskip .1cm
\centerline{\it Institute for Fundamental Theory}
\centerline{\it    Department of Physics}
\centerline{\it    University of Florida, Gainesville, FL 32611, USA}
{\smallskip\noindent
We extend the Wolfenstein parametrization to the quark masses in the deep
ultraviolet and develop an algorithm to derive symmetric textures which are
compatible with existing data. We find there are only five such textures.}

\endpage
\centerline{I. {\bf INTRODUCTION}}
\vskip .4in

The remarkable success of the standard model of strong and electroweak
interactions can be traced to one fact: the forces among elementary particles
are generated by spin one particles, photons, $W$ and $Z$ bosons, and gluons.
We
are incredibly lucky because their renormalizable interactions must be of
Yang-Mills type, which severely and happily restricts their possible
interactions. In this case, symmetry imposes the dynamics.

However this is not true of all interactions predicted by the standard model.
Masses of the elementary fermions come from their interactions with spin zero
Higgs particle. These Yukawa interactions have so far escaped all attempts at
classification in terms of a recognizable symmetry principle. Yet we have a
substantial amount of information about their values, in the form of lepton
masses, quark masses and mixing angles. However,
the precise values of the masses of the
lighter quarks are not accurately known since the concept of quark mass does
not make sense except at very high energy, far out of experimental reach.

It is well known that  the similarity between the quantum numbers of quarks and
leptons led to the concept of Grand Unification; it suggests that at some very
small distance, the gauge couplings should be equal. We may hope for
a similar situation for Yukawa couplings: their symmetries may be recognizable
only at short distances comparable to that at which the gauge forces unify.
In this paper I report on work done in collaboration with R. G. Roberts and G.
G. Ross\refto{RRR}, on one particular aspect of this hunt: we use the
available experimental data to restrict allowed textures in the Yukawa
matrices.
\vskip .4in
\centerline{II. {\bf MODELS OF YUKAWA MATRICES}}
\vskip .4in

The search for a recognizable simplicity among the Yukawa couplings has led
researchers to consider various elegant possibilities, none of which have
proven
to be predictive. These ideas include models with family symmetries, both
discrete and continuous; models with perturbative hierarchies, and models based
on chiral symmetry breaking condensates.
In the standard model, the Yukawa part of the Lagrangean is given by

$${\cal L}_Y =  L_i^T \bar e_j H^\ast {\bf Y}_e^{*ij} + {\bf Q}_i^T
\bar {\bf d}_j H^\ast {\bf Y}_d^{*ij} +
 {\bf Q}_i^T \bar {\bf u}_j \tau_2 H {\bf Y}_u^{*ij} +
{\rm c.c.}\ ,$$
where $i,j=1,2,3$ are the family indices. The standard model $3\times 3$ family
matrices ${\bf Y}_u\ ,{\bf Y}_d\ ,{\bf Y}_e$ need not display any particular
symmetry in family space.  In order to endow these matrices with symmetry
characteristics, it is necessary to reach for theoretical structures beyond the
standard model.

In GUT extensions of the standard model, Yukawa matrices acquire particular
symmetry
characteristics, depending on the particular Higgs to which they couple:

\noindent$\bullet$ In $SU(5)$, the up quark matrix is family symmetric for the
{\bf 5} and {\bf 50} Higgs representations, antisymmetric for the {\bf 45}; the
lepton and down quark matrices are indeterminate.

\noindent$\bullet$ In $SO(10)$,
all the Yukawa matrices have definite symmetry or antisymmetry, depending on
the
Higgs representation. All three Yukawa matrices merge into one; it is symmetric
for the ${\bf 10}$ and ${\bf 126}$ representation, antisymmetric for the ${\bf
120}$.

\noindent$\bullet$ In $E_6$, we have a similar situation to that in $SO(10)$:
the symmetric Yukawas correspond to Higgs in the
${\bf 27}$ and ${\bf 351'}$, the
antisymmetric in the ${\bf 351}$. However, the simplest Higgs representations
lie in the family-symmetric sector.

In addition, GUTs imply an interesting pattern of mass relations, namely
relation between lepton and charge -1/3 quark masses (in the so-called flipped
$SU(5)$, this association is lost). Assuming parcimony in the Higgs fields,
this
has led to the mass relation\refto{CEG}

$$m_b=m_\tau\ ;$$
Similar relations do not work for the lighter families, but are replaced
by\refto{GJ}

$$m_e={1\over 3}m_d\ ;\qquad m_\mu=3m_s\ .$$
While GUTs allow for succesful relations between lepton and down quark masses,
they do not provide any clues to relate masses of quarks of different charges,
such as the ratio of the bottom to top quark masses, nor to the relative values
of the quark masses within the same charge sector, such as the ratio of the
down
to strange quark masses, etc... . Thus, even with GUTs, there are two distinct
hierarchies to explain--the hierarchy among particles of the same charge, and
the hierarchy between particles of different charges. In the following, we
limit
our discussion to the hierarchy between particles of the same charge.

The most remarkable thing about the Yukawa coupling matrices is that the third
chiral family is so much more massive than the lighter two. This suggests that
all three Yukawa matrices are approximately proportional to\refto{GMRS}

$$\pmatrix{0&0&0\ccr 0&0&0\ccr
0&0&1\ccr}\ ,$$
which implies an approximate chiral symmetry $SU(2)_L\times SU(2)_R$ in all
three charge sectors. Thus one searches for theories where the form
of this matrix is natural, {\bf and} where there are small parameters to fill
in
the zeros. In addition, the hierarchy of masses of the particles of the lighter
families suggests we seek a perturbation\refto{CALC} in which the second family
gets its
mass in first order and the lightest family becomes massive in second
order\refto{INV}.

One set of theories has introduced  vectorial global family symmetries, both
discrete and continuous. Under an $SO(3)$ family group\refto{WZ},
the zeros in the 11 and 22 elements of the zeroth order matrix would result
from
an exact cancellation between the quadrupole and the singlet representation,
making it somewhat unatural. Thus it is better to assume an
$SU(2)$\refto{MSPIN}
or $SU(3)$\refto{PMR} family structure. The $SU(2)$ case naturally allows for
one zero eigenvalue, since the quantum number structure is

$$\pmatrix{{\bf 2}&{\bf
1}\ccr}\pmatrix{{\bf 3}&{\bf 2}\ccr {\bf 2}&{\bf 1}\ccr}\pmatrix{{\bf
2}\ccr{\bf
1}\ccr}\ ,$$
which allows no entries if the Higgs sector contains only family
doublets. The $SU(3)$ case is more natural since the zeroth order matrix now
transforms as a sextet.

There have been other attempts at a natural explanation of this hierarchy,
based
on the perturbative structure of field theory, but they have yielded theories
with a greater degree of complication.

In strong coupling models, the zeroth order Yukawa matrix
suggests a universal mechanism by which the chiral symmetry is
broken, because it is unitarily equivalent to the {\it democratic} matrix
which has all its entries equal.\refto{DEMOS}

In this paper we do not present any new theoretical insight into these
questions, but rather discuss the ways in which the low energy data on these
matrices might be used to divine the underlying cause of that structure.
\vskip .4in

\centerline{III. {\bf QUARK MASSES AND MIXING ANGLES}}
\vskip .4in

The observables in the Yukawa sector of the standard model are the fermion
masses and the elements of the Cabibbo-Kobayashi-Maskawa (CKM) matrix.
All but the top quark masses are known. The lighter quark masses are extracted
from the effective chiral model to be\refto{LEUT}

$$m_u=5.2~ \MeV\ ;~m_d=9.2~\MeV\ ;~m_s=194~\MeV\ ;\ {\rm all~~at~~1~~GeV}$$

\noindent The heavier quark masses have physical values

$$M_c=1.54\pm .04~ \GeV\ ; M_b=4.89\pm .04~ \GeV \ ,$$

\noindent corresponding to the running values at their respective physical
masses

$$m_c(M_c)=1.22 \pm .06~ \GeV~;~~m_b(M_b)=4.32\pm .06~ \GeV \ .$$

The top quark is known (assuming the standard model) to have a mass between 120
(my unofficial estimate from the lack of clear events at D0 and CDF), and 200
GeV from the Veltman bound\refto{VELT} (the ratio of neutral to
charged current processes).
Ratios of light quark masses are not expected to run much with scale, but those
that involve the heavier quarks, especially the top quark, run significantly
because of the Pendleton-Ross fixed point\refto{PEND}.

As mentioned in the introduction, it might be more interesting to extrapolate
these values in the deep ultraviolet where the theory could be simpler, as
suggested by GUTs. For this, we need to make certain assumption about physics
at
shorter distances. We consider two cases, that of no new physics ($N=0$), and
that of
low energy supersymmetry ($N=1$).

We first extrapolate these masses to the deep ultraviolet, using only the
standard
model. In the table below, we show the rough numerical values\refto{ARA} of the
quark mass
ratios at $10^{15}$ GeV in the $(N=0)$ standard model.
\vskip 1cm
\begintable
\multispan 7 \tstrut \hfill $N=0$ Standard Model: $10^{15}$ GeV  \hfill
\crthick
--- | ${m_u\over m_c}$ | ${m_c\over m_t}$ | ${m_u\over m_t}$
| ${m_d\over m_s}$ | ${m_s\over m_b}$ | ${m_d\over m_b}$  \crthick
$M_t=100$ GeV | .0037 | .0070 | .000025 | .047 | .032 | .0015 \cr
$M_t=150$ GeV | .0037 | .0043 | .000016 | .047 | .033 | .0016 \cr
$M_t=190$ GeV | .0036 | .0030 | .000009 | .047 | .036 | .0017 \endtable
\vskip .3in
In the minimal $(N=1)$ super-standard model, with the average threshold of
supersymmetry at $1$ TeV, these values change to
\vskip .3in
\begintable
\multispan 7 \tstrut \hfill $N=1$ Standard Model: $10^{15}$ GeV  \hfill
\crthick
$M_{SUSY}=$ 1 TeV | ${m_u\over m_c}$ | ${m_c\over m_t}$ | ${m_u\over m_t}$ |
${m_d\over m_s}$ |  ${m_s\over m_b}$ | ${m_d\over m_b}$  \crthick
$M_t=100$ GeV | .0037 | .0068 | .000025 | .047 | .030 | .0014  \cr
$M_t=150$ GeV | .0036 | .0038 | .000009 | .047 | .029 | .0014\cr
$M_t=190$ GeV | .0036 | .0022 | .000004 | .047 | .026 | .0012
\endtable
\vskip .3in
The purpose of the tables below is to give an idea of the magnitudes involved,
keeping in mind the uncertainties associated with the light quark masses.
These tables can lead to some intriguing numerology; for instance, with
a heavy top quark mass in the $N=1$ case, we note the relation

$${m_c\over m_t}\approx \left({m_d\over m_s}\right)^2\ .$$

\centerline{IV. {\bf WOLFENSTEIN PARAMETRIZATION}}
\vskip .4in
The sum total of known experimental information about the CKM matrix can best
be
summarized in terms of the Wolfenstein parametrization \refto{WOLF}.
It is just an expansion in the Cabibbo angle $\lambda $, namely

$${\cal U}_{CKM}^W=e^{iA_4\mib\lambda_4}e^{iA_5\mib\lambda_5}
e^{iA_7\mib\lambda_7}e^{iA_2\mib\lambda_2}\ ,$$
\noindent where the $\mib\lambda_i$ are the Gell-Mann matrices, and the
parameters are

$$A_2=\lambda \ ;~A_7=A\lambda ^2\ ;~A_5=A\rho \lambda ^3\ ;
{}~A_4=A\eta\lambda^3\ .$$

\noindent The first three parametrize the rotations, and $A\eta$ denotes the CP
violating phase. In matrix form, it is given by

$${\cal U}_{CKM}^W= \pmatrix{1-{\lambda^2\over 2}&\lambda
&A\lambda^3(\rho+i\eta)
\ccr-\lambda &1-{\lambda ^2\over
2}&A\lambda ^2\ccr
A\lambda ^3(1-\rho+i\eta)&-A\lambda ^2&1\ccr}\ ,$$

\noindent where $A,~\rho,$ and $\eta$ are of order one.
The latest experimental information yields the following values for its
parameters\refto{BURAS}

$$\lambda\approx .221\pm.003\ ;\ A\approx 0.85\pm0.09\ .$$
\noindent The other two parameters are not precisely known because of our
ignorance of the value of $f_B$. We have two rough ranges
$$\rho+i\eta=\cases{+0.3+i0.42\ ,\ f_B\approx 225\  \MeV \ccr\noalign{\vskip
.2cm} -0.4+i0.25\ ,\ f_B\approx 130\ \MeV\ccr} \ ,$$ with errors of the order
of
the values themselves. The uncertainties in these parameters are mostly
theoretical, due to our lack of knowledge of the B decay constant.

Our new input to this parametrization is to extend it to the quark masses, not
at experimental scales\refto{FISH}, but in the deep ultraviolet; we set
accordingly

$$\eqalign{{m_u\over m_t}&\equiv \mu_{u} \lambda ^8\ ;\qquad{m_c\over m_t}
\equiv \mu_{c}\lambda^4\ ;\ccr\noalign{\vskip .3cm}
{m_d\over m_b} &\equiv \mu_{d} \lambda^4\ ;\qquad{m_s\over m_b}\equiv
\mu_{s} \lambda ^2 \ .\ccr\noalign{\vskip .2cm}}$$
The parameters $\mu_{u},\mu_{c},\mu_{d},\mu_{s}$ are of order one, and
generally complex.
Their phases are not physical since they can be absorbed by redefining the
right-handed quark fields. Using the renormalization group to extrapolate into
the deep ultraviolet, we can rewrite the tables above in terms of the
magnitudes
of these ratios, in both the standard model and for its minimal supersymmetric
extension:
\vskip .4in
\begintable
\multispan 5 \tstrut \hfill $N=0$ Standard Model: $10^{15}$ GeV  \hfill
\crthick
--- | $\mu_{u}$ | $\mu_{c}$ | $\mu_{d}$ | $\mu_{s}$  \crthick
$M_t=100$ GeV | 4.72 | 3.06 | 0.65 | 0.68   \cr
$M_t=150$ GeV | 3.02 | 1.89 | 0.69 | 0.70 \cr
$M_t=190$ GeV | 1.70 | 1.31 | 0.74 | 0.75
\endtable
\vskip .5in
\begintable
\multispan 5 \tstrut \hfill $N=1$ Standard Model: $10^{15}$ GeV  \hfill
\crthick
$M_{SUSY}=1$ TeV | $\mu_{u}$ | $\mu_{c}$ | $\mu_{d}$ | $\mu_{s}$ \crthick
$M_t=100$ GeV | 4.72 | 2.95 | 0.61 | 0.62 \cr
$M_t=150$ GeV | 1.70 | 1.65 | 0.61 | 0.60 \cr
$M_t=190$ GeV | 0.75 | 0.95 | 0.52 | 0.54
\endtable
\vskip .4in
Note that for a heavy top quark, the $N=1$ case yields almost nice values,
close
to rational numbers. For a $``$light" top quark, this parametrization is not as
good, since $\mu_{u}\lambda$ can be of order one.
Still we will find it convenient in determining the validity of possible
textures in the Yukawa coupling constants.

\vskip .4in
\centerline{V. {\bf QUARK YUKAWA MATRICES}}
\vskip .4in

In the following, we assume that the quark Yukawa matrices are symmetric in
family space. Symmetric matrices can be diagonalized by means of a Schur
transformation. The up matrix can be written in the form,

$${\bf Y}_u={\bf V}_u{\bf M}_u{\bf V}^T_u\ ,$$

\noindent where ${\bf V}_u$ is a unitary matrix, and ${\bf M}_u$ is a diagonal
matrix with entries given by the masses $m_u\ ,m_c\ ,m_t$,

$${\bf M}_u=m_t\pmatrix{\mu_{u}~\lambda ^8&0&0\ccr 0&\mu_{c}~\lambda ^4&0\ccr
0&0&1\ccr}\ .$$
\noindent Note that the $\mu$ parameters are allowed to be complex; this
entails
no physical consequences since the phases can be absorbed in the right-handed
quark fields.

A similar transformation, applied to the down matrix, yields the result

$${\bf Y}_d={\bf V}_d{\bf M}_d{\bf V}^T_d\ ,$$

\noindent where ${\bf V}_d$ is unitary, and ${\bf M}_d$ diagonal with $m_d\
,m_s\
,m_b$ as entries

$${\bf M}_d=m_b\pmatrix{\mu_{d}~\lambda ^4&0&0\ccr 0&\mu_{s}~\lambda ^2&0\ccr
0&0&1\ccr}\ .$$

\noindent  The observable CKM matrix is given in terms of ${\bf V}_u\ ,{\bf
V}_d$ by

$${\cal U}_{CKM}={\bf V}_u^\dagger{\bf V}_d\ ,$$

\noindent  which enables us to eliminate ${\bf V}_d$ and
rewrite ${\bf Y}_d$ as

$${\bf Y}_d={\bf V}_u{\cal U}_{CKM}~{\bf M}_d~{\cal U}_{CKM}^T{\bf V}_d^T\
.$$

In order to sort out the phases, it is convenient to decompose the unitary
matrices in their {\it Iwasawa} form, which is a generalization of the Euler
decomposition. In particular, the CKM matrix is related to its Wolfenstein
parametrization through the formula

$${\cal U}_{CKM}={\cal P}{\cal U}_{CKM}^W{\cal P}'\ ,$$

\noindent where ${\cal P}$ and ${\cal P}'$ are diagonal matrices with pure
phase
as elements.
We note that the inner phase matrix ${\cal P}'$ is not relevant, as it
can be absorbed into the down quark masses, and then transferred to the
right-handed fields. It proves convenient to introduce the symmetric matrix

$${\bf X}_d\equiv{\cal U}_{CKM}^W~{\bf M}_d~{\cal U}_{CKM}^{WT}\ ,$$

\noindent which is simply expressed in terms of the observable Wolfenstein
parameters and down quark masses. To the lowest order in $\lambda$, it is given
by

$${\bf X}_d=m_b\pmatrix{(\mu_{s}+\mu_{d})\lambda ^4&\mu_{s} \lambda ^3&A(\rho
+i\eta)\lambda^3\ccr
\mu_{s} \lambda ^3&\mu_{s} \lambda ^2&A \lambda ^2\ccr
A(\rho +i\eta)\lambda ^3&A \lambda ^2&1\ccr}\ .$$

The top and bottom quark Yukawa matrices are now expressed in terms of the
symmetric matrix
${\bf X}_d$, the phase matrix ${\cal P}$, and the unitary matrix ${\bf V}_u$

$$\eqalign{{\bf Y}_u&={\bf V}_u{\bf M}_u{\bf V}^T_u\ ,\ccr
{\bf Y}_d&={\bf V}_u{\cal P}{\bf X}_d{\cal P}{\bf V}^T_u\ ;\ccr}$$

We can further simplify these by weeding out more extraneous phases. The
unitary matrix that provides the Schur decomposition of the Yukawa matrices can
itself be expanded {\it\` a la Iwasawa},
$${\bf V}_u={\cal P}_u{\bf\hat V}_u{\cal P}_u'\ ,$$
where the inner matrix depends on three rotation angles and one phase.
This reflects the inherent phase ambiguities in the Schur decomposition.
We get

$$\eqalign{{\bf Y}_u&={\cal P}_u{\bf\hat V}_u
{\cal P}_u'{\bf M}_u{\cal P}_u'{\bf\hat V}^T_u{\cal P}_u\ ,\ccr
{\bf Y}_d&={\cal P}_u{\bf\hat V}_u{\cal P}_u'{\cal P}{\bf X}_d
{\cal P}{\cal P}_u'{\bf\hat V}^T_u{\cal P}_u\ ;\ccr}$$

In this form, the phase ambiguities are made explicit.
We first absorb ${\cal P}_u'$ in the up quark masses, and then redefine the
right-handed up quark fields accordingly. This leaves us with the two phases
in ${\cal P}$. The overall phase is irrelevant. We can understand these two
phases to be equivalent to a redefinition of the parameters in the matrix
${\bf X}_d$, as follows
$$\mu_{d,s}\rightarrow \mu_{d,s}e^{i\phi_\mu}\ ;\qquad\lambda\rightarrow\lambda
e^{i\phi_\lambda}\ ;\qquad A\rightarrow  Ae^{i({\phi_\mu\over
2}-\phi_\lambda)}\
.$$
In the following we will assume these parameters to be complex with their
phases shown above; this is equivalent to absorbing the phase matrix ${\cal
P}$.
Thus, by specifying the unitary matrix ${\bf\hat V}_u$, and the phase matrix
${\cal P}_u$, we determine the Yukawa quark matrices, assuming that ${\bf X}_d$
is known.

$$\eqalign{{\bf Y}_u&={\cal P}_u{\bf\hat V}_u
{\bf M}_u{\bf\hat V}^T_u{\cal P}_u\ ,\ccr
{\bf Y}_d&={\cal P}_u{\bf\hat V}_u{\bf X}_d{\bf\hat V}^T_u{\cal P}_u\ ;\ccr}$$

These equations form the starting point of our computational algorithm
to be described in the next section.

This choice of input matrices is not unique;
we could have chosen to use the CKM matrix to eliminate ${\bf V}_u$.
This leads to defining the observable symmetric matrix

$${\bf X}_u\equiv{\cal U}_{CKM}^{W\dagger}~{\bf M}_u~{\cal U}_{CKM}^{W*}\ ;$$

\noindent in terms of the the Wolfenstein parameters  and the up quark mass
ratios, it is given by

$${\bf X}_u=m_t
\pmatrix{[\mu_{c}+A^2(1-\rho-i\eta)^2]\lambda
^6&[-\mu_{c}-A^2(1-\rho-i\eta)]\lambda ^5&
A(1-\rho-i\eta)\lambda ^3\ccr[-\mu_{c}-A^2(1-\rho-i\eta)]\lambda
^5&(\mu_{c}+A^2)\lambda ^4&-A\lambda ^2\ccr
A(1-\rho-i\eta)\lambda ^3&-A\lambda ^2&1\ccr}\ .$$

This time it is the phase matrix ${\cal P}'$ that is absorbed by the up quark
masses. It follows that the quark Yukawa matrices are given by

$$\eqalign{{\bf Y}_d&={\bf V}_d{\bf M}_d{\bf V}^T_d\ ,\ccr
{\bf Y}_u&={\bf V}_d{\cal P}^{*'}{\bf X}_u{\cal P}^{*'}{\bf V}^T_d\ . \ccr}$$

The {\it Iwasawa} decomposition is applied to the unitary matrix ${\bf V}_d$,

$${\bf V}_d={\cal P}_d{\bf\hat V}_d{\cal P}_d'\ ,$$
and we absorb the pases accordingly, to arrive at the final expressions

$$\eqalign{{\bf Y}_d&={\cal P}_d{\bf\hat V}_d{\bf M}_d{\bf\hat V}^T_d{\cal
P}_d\ ,\ccr
{\bf Y}_u&={\cal P}_d{\bf\hat V}_d{\bf X}_u{\bf\hat V}^T_d
{\cal P}_d\ . \ccr}$$

As above, the parameters in the matrix ${\bf X}_u$ are taken to be complex,
with
the phase of $A$ twice that of $\mu_c$.
This form is convenient for a simple down Yukawa matrix; the
unitary matrix ${\bf\hat V}_d$, and one phase matrix need to be specified.
\vskip .4in

\centerline{VI. {\bf ALGORITHMIC SEARCH FOR YUKAWA QUILTS}}
\vskip .4in
We see from the above that we have reduced the catalog of possible Yukawas to
that of possible unitary matrices ${\bf\hat V}_u$
(or ${\bf\hat V}_d$, depending on the starting point).
Let us start with ${\bf X}_d$ and ${\bf\hat V}_u$ as input matrices.
\bigbreak
The unitary matrix ${\bf V}_u$ contains eight parameters, and one overall phase
of no physical import. Its parameters need not be related to the observable
parameters contained in ${\bf M}_u$ and ${\bf X}_d$, unless the
Yukawa matrix elements obey special relations. What can those be? One piece of
evidence was the experimentally successfull relation between the Cabibbo angle
and the ratio
of down to strange quark masses, the Gatto, Sartori, Tonin, and Oakes
(GSTO)\refto{GSTO} relation,
$$\lambda\approx\sqrt{m_d\over m_s}\ ,$$
which relates a mixing angle to a ratio of eigenvalues. It was noted by
Weinberg, and Wilczek and Zee\refto{SW},
that a symmetric $2\times 2$ matrix with one zero diagonal element is
diagonalized by a rotation whose angle is related to the ratio of its
eigenvalues in precisely that way. An early generalization to the
$3\times 3$ case was proposed by Fritzsch\refto{HF}.
Other $``$textures" soon followed,
based on GUTs, by Georgi and Jarlskog\refto{GJ}, by Harvey, Ramond,
Reiss\refto{HRR}. More recently,
the GJ-HRR texture was analyzed in the context of a supersymmetric theory by
Dimopoulos, Hall, Raby,\refto{DHR} and by the Florida group\refto{PIERRE,ACPR}
. Another
interesting texture has been proposed by Giudice\refto{GIU}.
However there has been no systematic study of the
possible textures allowed by experiment. It is the
purpose of this paper to catalog the possible textures of Yukawa matrices which
are experimentally tenable. Each texture carries with it predictions among
experimental observables.

We note that the GSTO relation is obtained by punching a zero in the 11
position
of the embryonic Yukawa matrix ${\bf X}_d$. Similarly, punching a zero in the
22
position of the embryonic Yukawa matrix ${\bf X}_u$ yields the HRR
relation\refto{HRR}
$$V_{cb}=\sqrt{m_c\over m_t}\ ;$$
With the present experimental value of $V_{cb}$, it predicts a top quark mass
around 190 GeV. There are still large uncertainties, mostly theoretical, in the
extraction of $V_{cb}$ from experiment.

The careful analysis of the phases, carried out in the previous section, shows
that not all of the eight parameters in ${\bf V}_u$ are relevant. Indeed, the
relations

$$\eqalign{{\bf Y}_u&={\cal P}_u{\bf\hat V}_u{\bf M}_u{\bf\hat V}^T_u{\cal
P}_u\ ,\ccr
{\bf Y}_d&={\cal P}_u{\bf\hat V}_u{\bf X}_d{\bf\hat V}^T_u{\cal P}_u\ .\ccr}$$

The outer phase matrix ${\cal P}_u$ has no effect on the location of possible
zeros in the Yukawa matrices; we can therefore neglect it , and the search for
symmetric textures is reduced to specifying allowed forms of the reduced
unitary
matrix ${\bf\hat V}_u$, which we parametrize in the form

$${\bf\hat V}_u=e^{i\varphi\mib\lambda_4}{\cal R}(\theta_1,\theta_2,\theta_3)\
,$$

\noindent where the $\theta_i$ are the Euler rotation  parameters and $\varphi$
denotes the one (CP violating) phase. It is a simplifying feature of our method
of searching for zeroes in the Yukawa matrices that it only depends on these
four parameters, one of which is a phase. We have taken the phase to be along
the $\mib\lambda_4$ direction, to match with the Wolfenstein parametrization.
This choice proves convenient in our analysis. Having the phase in a different
direction corresponds to a different way of choosing the parameters.

To summarize, the hunt for zeros in the Yukawas starts
with the up quarks Yukawa written in the form

$${\bf Y}_u={\bf\hat V}_u{\bf M}_u{\bf\hat V}^T_u\ .$$
We can proceed in two distinct ways:

$\bullet$ one is to input the above parametrization
for the unitary matrix ${\bf\hat V}_u$ (there are
sixteen possibilities in order of increasing complexity), then
compute ${\bf Y}_u$, and determine which relations
can be enforced between parameters of ${\bf\hat V}_u$ and up quark mass ratios
by
punching zeroes in matrix elements of ${\bf Y}_u$; this enforces
a certain texture for the up quark Yukawa. We impose as many conditions as
there
are parameters in ${\bf\hat V}_u$.

$\bullet$ or we can assume a possible texture for ${\bf Y}_u$ consistent with
the hierarchy of quark masses (there are six of them), diagonalize the up
Yukawa
to obtain ${\bf\hat V}_u$.

Either procedure leaves us with a ${\bf\hat V}_u$, whose parameters depend on
up quark mass ratios. We proceed to compute the down Yukawa matrix according to

$${\bf Y}_d={\bf\hat V}_u{\bf X}_d{\bf\hat V}^T_u\ .$$
We then examine which of its matrix elements can be set equal to zero in a way
that is consistent with experimental constraints. Punching zeroes in the down
Yukawa will result in relations which involve the mixing angles as well as the
up and down quark mass ratios. In this way, for a given ${\bf\hat V}_u$,  we
obtain a particular texture for the up and down Yukawa matrices, together with
its concomitant set of relations among observable parameters.

In this way we obtain all possible symmetric textures consistent with present
day experimental constraints. Some leave us with predictions which can be
tested
by experiment. We believe these textures to be valid in the deep ultaviolet
where the theory has a chance of being simpler. Thus, comparison with
experiment
depends on extrapolations to experimental scales. The only possible
perturbative
extrapolation is that which uses low energy supersymmetry, which is our
favorite
scenario\refto{RR}. We refer the reader to our paper\refto{RRR} for detailed
comparisons:

\vskip .3in
$${\bf Y}_u=\pmatrix{0 & \sqrt{2}\lambda^6 & 0 \ccr
\sqrt{2}\lambda^6 & \lambda^4 & 0 \ccr
0 & 0 & 1\ccr }\ ,\qquad {\bf Y}_d= \pmatrix{0 & 2\lambda^4 & 0 \ccr
2\lambda^4 & 2\lambda^3 & 4\lambda^3 \ccr
0 & 4\lambda^3 & 1\ccr }\ ;$$

$${\bf Y}_u=\pmatrix{0 & \lambda^6 & 0 \ccr
\lambda^6 & 0 & \lambda^2 \ccr
0 & \lambda^2 & 1\ccr}\ ;\qquad {\bf Y}_d=
\pmatrix{0 & 2\lambda^4 & 0 \ccr
2\lambda^4 & 2\lambda^3 & 2\lambda^3 \ccr
0 & 2\lambda^3 & 1\ccr}\ ;$$

$${\bf Y}_u=\pmatrix{0 & 0 & \sqrt{2}\lambda^4 \ccr
0 & \lambda^4 & 0 \ccr
\sqrt{2}\lambda^4 & 0 & 1\ccr}\ ;\qquad
{\bf Y}_d=\pmatrix{0 & 2\lambda^4 & 0 \ccr
2\lambda^4 & 2\lambda^3 & 4\lambda^3 \ccr
0 & 4\lambda^3 & 1\ccr}\ ;$$

$${\bf Y}_u=\pmatrix{0 & \sqrt{2}\lambda^6 & 0 \ccr
\sqrt{2}\lambda^6 & \sqrt{3}\lambda^4 & \lambda^2 \ccr
0 & \lambda^2 & 1\ccr}\ ;\qquad {\bf Y}_d=\pmatrix{0 & 2\lambda^4 & 0 \ccr
2\lambda^4 & 2\lambda^3 & 0 \ccr
0 & 0 & 1\ccr}\ ;$$

$${\bf Y}_u=\pmatrix{0 & 0 & \lambda^4 \ccr
0 & \sqrt{2}\lambda^4 & \frac{\lambda^2}{\sqrt{2}} \ccr
\lambda^4 & \frac{\lambda^2}{\sqrt{2}} & 1\ccr}\ ;
\qquad {\bf Y}_d=\pmatrix{0 & 2\lambda^4 & 0 \ccr
2\lambda^4 & 2\lambda^3 & 0 \ccr
0 & 0 & 1\ccr}\ .$$

We have written these textures in a theoretically evocative way. We do not know
any deep reasons why such patterns should appear in the Yukawa couplings, but
we
might speculate. One possibility is that of discrete symmetries which are
abundant in string inspired theories\refto{GKR}. Another possibility, more
specifically connected to string theories, is that these Yukawa couplings may
be
of topological origin, describing intersection numbers on the coset manifolds
of
the theory. In this case the zeros would have natural explanations, denoting
lack of intersection between several curves on the manifold. We hope to return
to these issues at a later time.

I wish to thank Professors Kursunoglu and Perlmutter for inviting me to address
this conference.

\references

\refis{CEG}
M.S. Chanowitz, J. Ellis and M.K. Gaillard,  \np, B128, 506, 1977.
; A. Buras, J. Ellis, M.K. Gaillard and D.V. Nanopoulos, \np, B135, 66, 1978..

\refis{GJ} H. Georgi and C. Jarlskog, \pl, 86B, 297, 1979..

\refis{HRR} J. Harvey, P. Ramond and D. Reiss, \pl, 92B, 309, 1980.
; \np, B199, 223, 1982..

\refis{DHR}S. Dimopoulos, L.J. Hall and S. Raby, \prl, 68, 1984, 1992.
; \pr, 45D, 4195, 1992..

\refis{PIERRE} P.~Ramond, in ``On Klauder's Path : a Field
Trip'', eds. G. Emch, G. Hegerfeldt, L. Streit (World Scientific,
Singapore) 1992.

\refis{ACPR}
H. Arason, D.J. Casta\~no, P. Ramond and E.J. Piard, \pr, 47D, 232, 1993..

\refis{ARA}H. Arason, D.J. Casta\~no, B. Keszthelyi, S. Mikaelian,


\refis{SW} S. Weinberg, in `` A Festschrift for I.I.Rabi"
[Trans. N.Y. Acad. Sci., Ser. II (1977), v. 38], p. 185;
F. Wilczek and A. Zee, \pl, 70B, 418, 1977..

\refis{HF} H. Fritzsch, \pl, 70B, 436, 1977.; \pl, 73B, 317, 1978.
;  P. Kaus and S. Meshkov, \mpl, A3, 1251, 1988.
;  F.J. Gilman and Y. Nir, \arnp, 40, 213, 1990 ..


\refis{RRR} P. Ramond, R.G. Roberts, and G.G. Ross, RAL-93-010 and UFIFT-93-05.

\refis{FISH}P. Fishbane and P.Q. Hung, \pr, 45D, 293, 1992.
; Y. Koide, H. Fusaoka and C. Habe, \pr, 46D, R4813, 1992..


\refis{GSTO} R. Gatto, G. Sartori and M. Tonin, \pl, 28B, 128, 1968.
; R.J.~Oakes, \pl, 29B, 683, 1969.
; \pl, 31B, 620(E), 1970.
; \pl, 30B, 262, 1970..

\refis{RR} G.G.~Ross and R.G. Roberts, \np, 377B, 571, 1992.

\refis{BURAS} A.~Buras and M.K.~Harlander, Munich preprint MPI-
PAE/PTh 1/92;
J.L.~Rosner, Jour. Phys. {\bf G18}, 1575(1992) .

\refis{GKR} B.R.~Greene, K.H.~Kirklin, P.J.~Miron, and G.G.~Ross,
\np, 278B, 667, 1986..


\refis{GMRS}M. Gell-Mann, P. Ramond, and R. Slansky as reported by P. Ramond in
in Sanibel Talk, CALT-68-709, Feb 1979, and in {\it Supergravity} (North
Holland, Amsterdam 1979).

\refis{MSPIN}The $SU(2)$ family group was introduced long ago under the name of
M-spin. See for instance F. G\" ursey and G. Feinberg, \pr, 128, 378, 1962.
; T.D. Lee, \nc, 35, 975, 1965.
; S. Meshkov and P. Rosen, \prl, 29, 1764, 1972.
; {\it ibid} \prd, 10, 3520, 1974..

\refis{PMR}P. Ramond, CALT-68-709, Feb 1979 (unpublished).

\refis{LEUT}J. Gasser and H. Leutwyler, {\it Phys. Rep.} {\bf 87}, 77(1982).

\refis{WZ}F. Wilczek, and A. Zee, \prl, 42, 421, 1979..

\refis{WOLF} L.Wolfenstein, \prl, 51, 1945, 1983..

\refis{VELT} M. Veltman, \np, B123, 89, 1977..

\refis{PEND} B. Pendleton and G.G. Ross, \pl, 98B, 291, 1981.
; C. Hill, \pr, D24, 691, 1981..

\refis{GIU}G. F. Giudice, \mpl, A7, 2429, 1992. .

\refis{DEMOS}H. Harari, H. Haut, and J. Weyers, \pl, 78B, 459, 1978.;
Y. Koide, \pr, D28, 252, 1983.;
P. Kaus and S. Meshkov, \mpl, A3, 1251, 1988..

\refis{CALC} S. Weinberg, \pr, D5, 1962, 1972.;\prl, 29, 388, 1972.;
H. Georgi and S. L. Glashow, \pr, D6, 2977, 1972.; \pr, D7, 2457, 1973.;
R. N. Mohapatra, \pr, D9, 3461, 1974.;
S. Barr, \pr, D24, 1895, 1981.; R. Barbieri and D. Nanopoulos, \pl, 91B, 369,
1980.; M. Bowick and P. Ramond, \pl, 103B, 338, 1981.;
R. Barbieri, D. Nanopoulos and A. Masiero, \pl, 104B, 194, 1981.;
R. Barbieri, D. Nanopoulos and D. Wyler, \pl, 106B, 303, 1981..

\refis{INV} For an intriguing twist, see Z. Berezhiani and R. Rattazi,
LBL-32889, Nov 1982.

\endreferences

\end